\documentclass{elsart}
\usepackage{natbib}   
\usepackage{graphicx} 
\begin{document}

\runauthor{Owsianik et al.}

\begin{frontmatter} 
\title{The Youngest Lobe-Dominated Radio Sources} 
\author[TCfA,MPIfR]{I. Owsianik\thanksref{TMR}}
\author[OSO]{J.E. Conway}
\author[OSO,JIVE]{A.G. Polatidis}  

\thanks[TMR]{I.O acknowledges financial support from OSO and JIVE}

\address[TCfA]{Toru\'n Centre for Astronomy, ul. Gagarina 11, 87-100
Toru\'n, Poland}
\address[MPIfR]{Max-Planck-Institute f\"ur Radioastronomie,
Auf dem H\"ugel 69, D-53121 Bonn, Germany}
\address[OSO]{Onsala Space Observatory, S-43992 Onsala, Sweden}
\address[JIVE]{Joint Institute for VLBI in Europe, Postbus 2, 
7990~AA Dwingeloo, The Netherlands}

\begin{abstract} 

We present an analysis of multi-epoch global VLBI observations
of the Compact Symmetric Objects: 2352+495 and 0710+439 at 5 GHz.  
Analysis of
data spread over almost two  decades shows strong evidence for an increase in
separation of the outer components of both sources 
at a rate of $\sim 0.2 h^{-1}$c
(for  $q_{\circ}$=0.5   and $H_{\circ}=100 h$  kms$^{-1}$Mpc$^{-1}$).
Dividing the overall sizes of the sources by their separation rates
implies that these Compact Symmetric Objects have 
 a kinematic age $\ll 10^{4}$ years. These results (and those for other CSOs)
strongly argue  that CSOs are indeed very young 
sources and that they are probably evolve into the much larger
classical doubles.

\end{abstract}


\end{frontmatter} 

\section{Introduction}
\label{intro} 

There exists a class of  enigmatic sources in 
which high luminosity radio emission regions are located `symmetrically' on both 
sides of the central engine on linear scales of less than 1 kpc; the so-called 
Compact Symmetric Objects \citep{wpr94}.
Several theories have been suggested to explain CSOs.
It has been suggested that they are a young phase in the development
of classical doubles \citep[e.g.,][]{pm82,ffd95,rtp96} or
that they are fairly old sources in which a dense 
environment inhibits their growth \citep[e.g.,][]{vmh84}. 
Finally it has been proposed that they are 
a separate class of short lived objects which `fizzle out' after 
about $10^{4}$ years \citep{rxp94a}.
An obvious way to try to 
distinguish between these models is to investigate the growth in
overall sizes of CSOs and hence set limits to their ages directly.

\section{Data analysis for 0710+439} 
\label{0710+439}

The radio source 0710+439 \citep[$z=0.518$,][]{lzr96} has a high 
radio luminosity  \citep[$L_{5GHz}=5\times10^{26} h^{-2}$ W Hz$^{-1}$,][]{rtx96}. 
The radio flux density
is very weakly polarised and it is not time variable \citep{aah92}. 

 
\begin{figure*}[ht] 
\centering
\includegraphics[scale=0.35,angle=0]{owsfig1a.ps}
\includegraphics[bb=0 60 550 550,scale=0.39,angle=0]{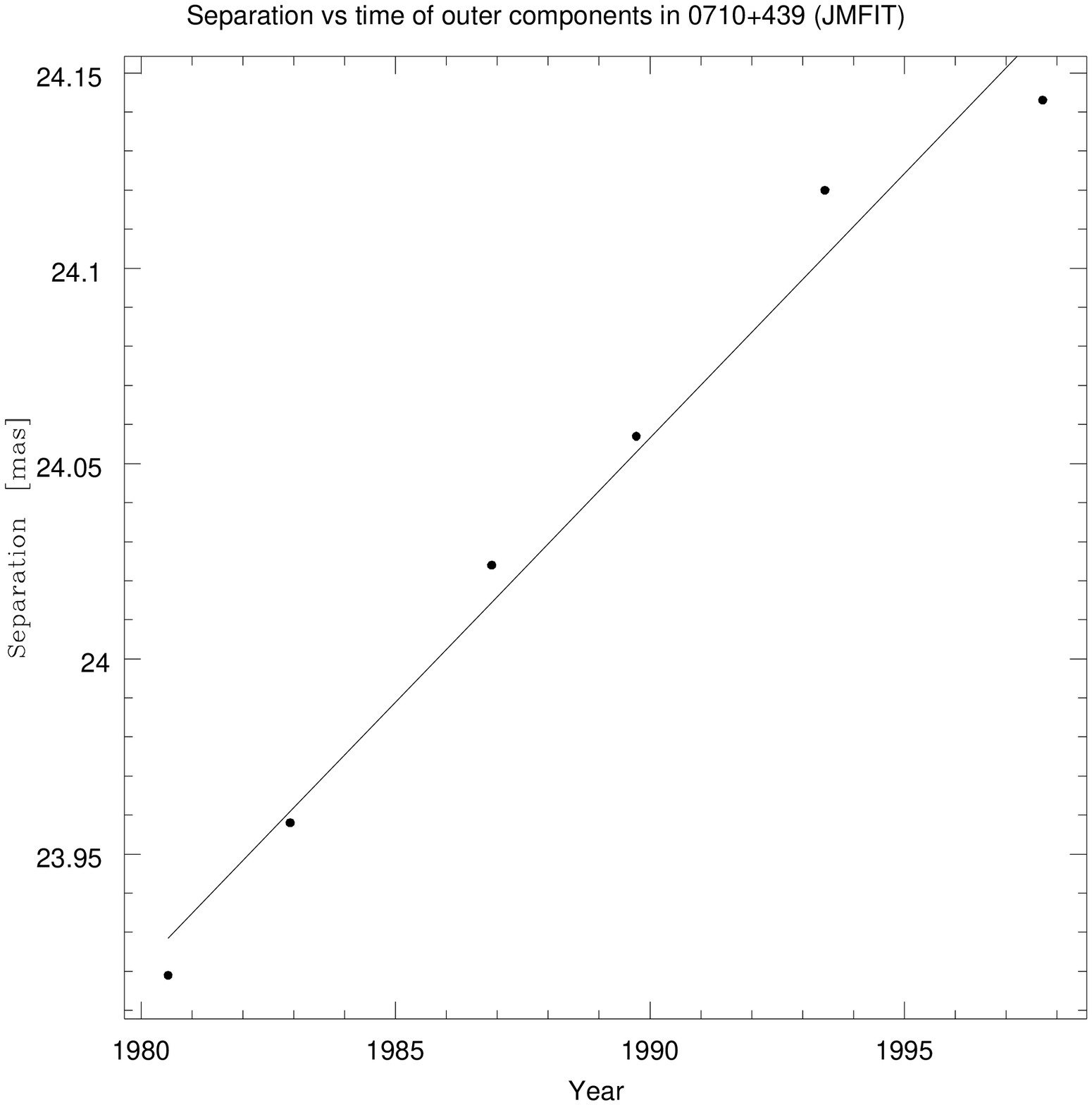}  
\caption{{\bf (a)} The natural weighted clean image of 0710+439 at 
5 GHz from the epoch 1997.71, rms noise   151 $\mu$Jy beam$^{-1}$,
{\bf (b)} Changes in separation with time between the hotspots. Dots represent data
obtained by JMFIT, solid line shows linear regression  fit. See text for a discussion
of the errors.}
\label{Fig_0710} 
\end{figure*}  


0710+439 has  been observed at   5 GHz with  a global  VLBI array at 6
epochs spread over  a period of 17  years. The first five epochs  were
analysed by \citet{oc98}.  Here we add new data from multi-snapshot
14 station  global observations made  at 18th  of  Sep 1997.
The highest  dynamic  range image (see Fig.  \ref{Fig_0710}a) was obtained
using DIFMAP  \citep{spt94} and  was used as
a starting point in re-mapping the other  epochs following the procedure 
describe by \citet{oc98}. This image shows clearly the overall triple structure
of the source. The northern and the  southern components show compact,
bright subcomponents which are associated with hotspots surrounded by the
faint extended lobe emission. The  middle component is associated with
the base of the  northern  jet, and the centre  of  activity lies at 
the southern  end of this component  \citep{trp96}.  The image in 
Fig. \ref{Fig_0710}a shows clearly for the first time 
at this frequency a bridge of emission  between northern hotspot and 
the  middle component.  In addition  
the emission previously detected  at 1.6 GHz \citep{x94} between the   
middle  and  southern components is for the first time now also detected at 5GHz.

The hotspot components are well separated on CLEAN images which allows us to use
the AIPS task JMFIT to fit the position of hotspot at each epoch. The fitted linear 
regression line to this data gives an estimated separation rate of
$13.614 \pm 0.988$ $\mu$as/yr (Fig. \ref{Fig_0710}b). From the obtained correlation coefficient  of 0.989 we can
reject the null hypothesis of no motion at a better than $0.1\%$  confidence level
\citep[for a fuller discussion on errors see][]{oc98}.
Dividing the distance between hotspots of 87.07 $h^{-1}$ pc by their 
observed separation velocity of $0.243 \pm 0.018 h^{-1}$c we estimate that  
the source 0710+439 is $1100 \pm 100$ years old.

Assuming that hotspots are close to their equipartition pressure and 
assuming that  the source is orientated not too far from the sky plane 
(as supported by the source having an arm-length ratio close to one) 
then ram pressure arguments  imply an external density of $2 h^{18/7}$ cm$^{-3}$. 
From the  velocity of advance of  0710+439 we can calculate that the rate of
work done advancing the two hotspots is $5 \times 10^{43} h^{-17/7}$ erg s$^{-1}$.
Following the arguments of \citet{rtx96} we can estimate an upper limit on the total jet power 
of $4 \times 10^{44} h^{-10/7}$ erg s$^{-1}$. A  lower limit on the jet  power 
 equals the sum of work of advance and radio luminosity ($7 \times 10^{43} h^{-2}$ erg s$^{-1}$).
Given these numbers we can estimate that  the efficiency ($\epsilon$)
of conversion  of jet mechanical energy  into radio emission in 0710+439
is  $23\% < \epsilon <53\%$ (for $h=0.6$).

\section{Data analysis for 2352+495}
\label{2352+495}

The radio source 2352+495  \citep[$z=0.238$,][]{lzr96} has a high radio luminosity
\citep[$L_{5GHz}=1\times10^{26} h^{-2}$ W Hz$^{-1}$,][]{rtx96}. The radio flux density
is very weakly polarised and it does not exhibit a significant time variation \citep{aah92}.
\citet{cpr92}  revealed for the first time the triple radio morphology of 2352+495
and from analysis of multi-epoch data it was argued that the center of activity  was 
associated with the middle component \citep{cpr92}. Later  15 GHz observations located 
a compact core component at the southern end of this middle component \citep{trp96}.

 
\begin{figure*}[ht] 
\centering
\includegraphics[scale=0.32,angle=0]{owsfig2a.ps}
\includegraphics[bb=0 100 590 450,scale=0.36,angle=0]{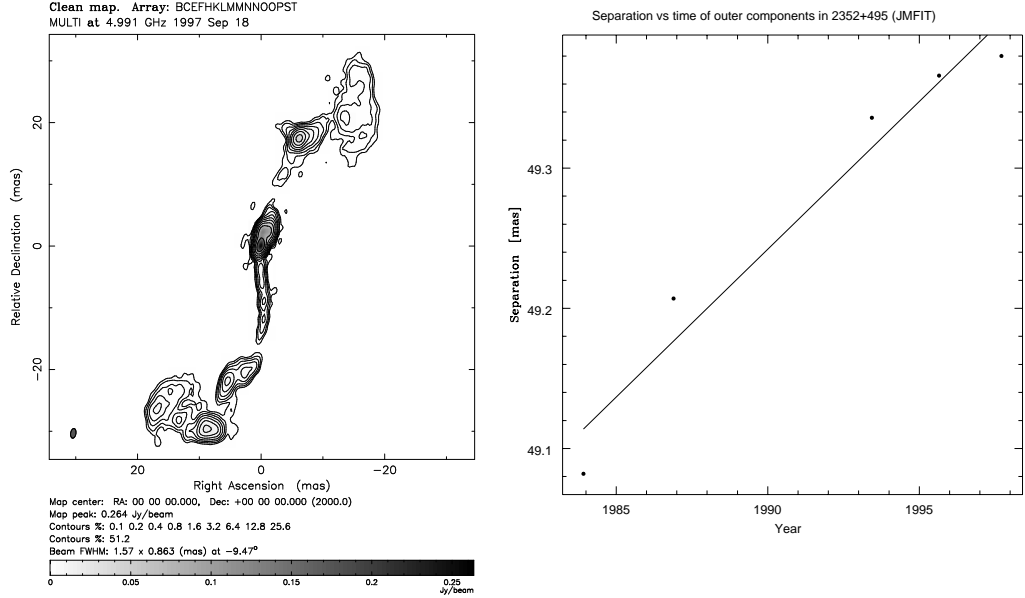} 
\caption{{\bf (a)} Natural weighted clean image of 2352+495 at 5 GHz from the epoch 1997.71, 
rms noise 57 $\mu Jy$ $beam^{-1}$,
{\bf (b)} Changes in separation with time between hotspots
obtained by JMFIT, solid line shows linear regression  fit. See text for a discussion
of the errors.}
\label{Fig_2352} 
\end{figure*}  


2352+495 has been observed at 5 GHz with a global VLBI 
array at five epochs spread over a period of 14 years. The
first two epochs  included in our analysis were made at epoch 1983.93  with a global
array of 6 telescopes and then at epoch 1986.89 using multiple snapshots 
with 9 telescopes \citep[see][]{cpr92}. Here we reanalyse these epochs and add data from
three additional epochs; a multi snapshot 12 station global VLBI observations 
made at epoch 1993.44, a 10 station VLBA observations made at epoch 1995.67 
(donated by G.B. Taylor and R.C. Vermuelen, gratefully acknowledged by the authors) and finally a 
14 station global VLBI observations made at epoch 1997.71.
The telescopes used included those from the European VLBI Network (EVN),
the Very Long Baseline Array (VLBA), the Very Large Array (VLA) and Haystack Observatory.

The  analysis of the data  followed the  procedure described
in \citet{oc98}.  Fig. \ref{Fig_2352}a shows the best image of 2352+495
obtained for the last epoch data  using DIFMAP \citep{spt94}, 
this map  was used as a starting point in analysis of the earlier epochs.  
This 5th epoch map clearly shows the overall triple
structure of the source. The northern and the southern components show
compact, bright subcomponents  which are associated with hotspots surrounded
by fainter extended  emission from   the lobes.  The middle  very bright
component  appears to be associated with the base of the  
northern jet \citep{trp96}. The image also shows an almost 
continuous  jet  connecting the middle component  with the
southern hotspot. For the first time this image  also 
reveals portions of a possible jet-like component on the other
side of the middle  component/core feature.

The well separated hotspot components in the CLEAN images  allowed us to 
use the AIPS task JMFIT to fit the position of hotspots at each epoch. 
The linear regression fits to the observed changes in gaussian  
component separation   gave us an estimate 
of the relative angular  separation rate of $21.062 \pm 2.704 \mu$as/yr corresponding  
to  a separation velocity of $0.202 \pm 0.026 h^{-1}$c (Fig. \ref{Fig_2352}b).
The linear regression fit gives a  correlation coefficient 
of 0.976 which allows us to reject the null hypothesis of no motion at
the  better than $1\%$  confidence level. Dividing the distance between hotspots 
of 117.03 $h^{-1}$ pc by their observed separation rate
we estimate that  2352+495 is $1900 \pm 250$ years old.

Assuming  that hotspots are close to  their equipartition pressure and
assuming  that  the source   is orientated  not   too far  from the  sky plane
(consistent with an arm-length ratio very close to 1)  then  ram pressure arguments imply
an external density of $1 h^{18/7}$ cm$^{-3}$,  which is consistent with
the NLR intercloud medium.  Given our  estimate of an  age and the jet
thrust  we can compare the  power  of advance   required to drive  the
hotspot forward,   with the   radio   luminosity  and the  jet   power
\citep{rtx96}.  For an  age of 2352+495  the rate of work done
in advancing the two hotspots is  $1 \times 10^{43} h^{-17/7}$  erg s$^{-1}$.
The upper limit on the   total power supplied by the jet is  $8  \times
10^{43} h^{-10/7}$ erg s$^{-1}$  \citep{rtx96}.  The lower  limit on
the total power of the jet is the sum of the power of advance
and the radio luminosity ($3 \times 10^{42} h^{-2}$ erg s$^{-1}$). Given these numbers we can 
therefore estimate that the efficiency of conversion of the  jet mechanical
energy into  the radio emission is $5\% < \epsilon < 19\%$ (for $h=0.6$).

\section{Evolution of CSOs}
\label{ages}

In addition to the two sources presented in this paper there are now
reliable velocities of expansion for three other CSOs; namely 
0108+388  \citep{ocp98}, 2021+614 (\citet{cmp94} and Tschager
et al., this volume) and 1943+456 (Polatidis, this volume). All
of the expansion velocities  are of order $0.2 h^{-1}$c; giving
hotspot advance speeds of an order $0.1 h^{-1}$c and  ages of 
a few thousand years. It has therefore become clear that 
the majority of CSOs  are very  young  sources, which  grow fairly   rapidly
\citep{oc98,ocp98}.
What is still uncertain is their subsequent evolution.
The  simplest scenario is that CSOs evolve via Medium-size Symmetric
Objects into Large-size Symmetric Objects
\citep{ffd95,rtp96,oc98}. Given the rapid expansion  rate of CSOs  the
sources will spend only a short time in the CSO phase, and it would be 
expected that only a small fraction of sources would be CSOs.  
In fact CSOs comprise about $7.5\%$ of sources in the flux   limited  surveys   
at 5GHz (Polatidis, this volume). However, this large fractional population can be
explained if, as is theoretically expected, there is a 
strong negative luminosity evolution  with
increasing  size. In  such a model   CSOs will evolve into
more        numerous,      but           much    weaker        sources
\citep[e.g.,][]{beg96,rtp96,oc98}.  As  noted by   \citet{rtx96}  for
2352+495, and  as we find here the  limits on radiative efficiency for
CSOs compared to `classical' large lobe-dominated
sources  \citep[in which $\epsilon$ is of  order of
few  per   cent][]{oc98} empirically  demonstrates  that   the expected
luminosity evolution does in fact  occur and with a size  consistent
with that expected  by theoretical models  \citep{beg96,rtx96}.

\end{document}